\documentclass[12pt,preprint]{aastex}

\slugcomment{to appear in the {\it Astronomical Journal}}

\shorttitle{White Dwarf-Red Subdwarf Systems}

\shortauthors{Monteiro et al.}

\begin{document}

\title{Ages of White Dwarf-Red Subdwarf Systems}

\author{Hektor Monteiro\altaffilmark{1}, Wei-Chun Jao\altaffilmark{1},
Todd Henry\altaffilmark{1}, John Subasavage\altaffilmark{1}, and Thom
Beaulieu\altaffilmark{1}}

\affil{Georgia State University, Atlanta, GA 30302-4106}

\altaffiltext{1}{Visiting Astronomer, Cerro Tololo Inter-American
Observatory.  CTIO is operated by AURA, Inc.\ under contract to the
National Science Foundation.}

\begin{abstract}

We provide the first age estimates for two recently discovered white
dwarf-red subdwarf systems, LHS 193AB and LHS 300AB.  These unusual
systems provide a new opportunity for linking the reliable age
estimates for the white dwarfs to the (measurable) metallicities of
the red subdwarfs.  We have obtained precise photometry in the
$V_{J}R_{KC}I_{KC}JH$ bands and spectroscopy covering from 6000\AA~to
9000\AA~for the two new systems, as well as for a comparison white
dwarf-main sequence red dwarf system, GJ 283 AB.  Using model grids
available in the literature, we estimate the cooling age as well as
temperature, surface gravity, mass, progenitor mass and {\it total}
lifetimes of the white dwarfs.  The results indicate that the two new
systems are probably ancient thick disk objects with ages of at least
6-9 Gyr.  We also conduct searches of red dwarf and white dwarf
compendia from SDSS data and the L{\'e}pine Shara Proper Motion (LSPM)
catalog for additional common proper motion white dwarf-red subdwarf
systems.  Only seven new candidate systems are found, which indicates
the rarity of these systems.

\end{abstract}


\section{Introduction}

Subdwarfs are Galactic fossils that provide answers to questions
concerning the age of the Milky Way, its early composition, the rate
of star formation at early epochs, and the Galaxy's overall
construction by the cannibalism of nearby small galaxies.  Subdwarfs
of K and M types are presumably the most numerous stellar components
of the early Galaxy, yet they are the types least studied because of
their low luminosities. Subdwarfs are notoriously far more rare than
disk stars, and therefore harder to identify and study. For example,
of the 243 stellar systems now known within 10 pc of the Sun, only two
are confirmed subdwarf systems (with accurate parallax and spectra):
$\mu$ Cas AB (GJ 53 AB, [Fe/H]$=-$0.71, \citealt[]{Karaali2003}) and
CF UMa (GJ 451, [m/H]$=-$1.44, \citealt[]{Carney1994})

Although they are difficult to find, a widely applied strategy used to
identify subdwarfs, selecting stars with high proper motions, can
yield a rich sample of subdwarf candidates that can be targeted for
more detailed observations. Selection by proper motion is highly
successful because old stars tend to have large heliocentric
velocities after billions of years of ``heating'' from many passages
through the Galactic disk. For example, Schmidt (1975), Gizis and Reid
(1999), and Digby (2003) have used high proper motion criteria to
select potential subdwarfs and to determine the subdwarf luminosity
function. However, we have found that only 19 (33\%) of Schmidt's 57
subdwarf candidates are, in fact, spectroscopically confirmed
subdwarfs \citep{Jao2004}, illustrating the significant contamination
of proper motion samples by disk stars.  Two observations are required
to scrub samples of potential low mass subdwarfs of interlopers: (1)
trigonometric parallax observations to confirm that candidates lie 1-3
magnitudes below the main sequence, and (2) spectroscopic observations
to confirm that candidates have low metallicity (strength of the CaH
and TiO bands in particular for the cool subdwarfs). Some subdwarf
candidates have been identified by one of these observations, but
within 60 pc there are currently only 58 K and M type subdwarfs
confirmed by both types of observations (Jao 2004).

White dwarfs (WDs) are the end products of the evolution of stars with
masses of 1 $M_{\odot}$ to 8 $M_{\odot}$. If we consider any typical
initial stellar mass function, it is instantly clear that this makes
WDs the most common end state of stellar evolution.  Because of their
large numbers, ability to be modeled effectively, and age dispersion
spanning our Galaxy's entire history, WDs provide important,
interpretable clues about Galactic timescales.  This potential was
recognized many years ago by \cite{S59} in a study of star formation
rates.  Many developments have occurred since then in both
observational and theoretical studies of WDs.  Observationally, with
modern telescopes we have been able to detect fainter objects, giving
us more detailed information on the cool, old WDs (\citealt{BLR01}),
including those in globular clusters that are used to determine
cluster ages \citep{Richer1997,Hansen2002}.  Recent observations also
suggest that an ancient WD population could exist in the Galactic halo
that may constitute an important part of the baryonic dark matter
\citep{Op01}, although this is still rigorously debated (see,
e.g. \citealt{Reid2001},
\citealt{RRC01},\citealt{DKR02},\citealt{SOH02},\citealt{Salim2004}
and \citealt{Bergeron2005}).  Theoretically, models (\citealt{BLR01},
\citealt{FBB01}) continue to improve our understanding of WDs, in
particular for the cool WDs that are members of the Galactic thick
disk or halo.

During our southern parallax program CTIOPI (Cerro Tololo Inter-American
Observatory Parallax Investigation), we have stumbled upon two unusual
WDs in systems with red subdwarfs, LHS 193AB and LHS 300AB\footnote{Both
systems have L447$-$10 and L395$-$13 identifiers, respectively, in the
New Luyten Two-Tenth (NLTT) catalogue. However, in order to keep consistent
with the IDs we used in \citet{Jao2003, Jao2005}, the LHS IDs will be
used throughout this manuscript.}  (Jao et al. 2005).  These two systems
contain the first WDs ever found to have confirmed subdwarf companions,
based on accurate parallax and spectroscopy, thereby giving us powerful
insight into the nature of old WDs in low metallicity systems.  Two
other systems like these have been mentioned in the literature, GJ 781AB
\citep{Gizis1998} and LHS 2139/2140 \citep{Gizis1997-2}, but both are
speculative, being based on broad spectral features plus mass functions
in unresolved binaries and noisy spectra of one component, respectively.
A third system, reported in \citet{Silvestri2002}, has been
spectroscopically confirmed but no parallax is available.  Table 1 lists
all of these systems as well as new candidate systems discussed in
section 5.  In this work we present a detailed analysis of the only two
such resolved systems with parallaxes, including photometric and
spectroscopic observations and estimates of the WDs' physical
parameters.  Using models from the literature, we then use the physical
parameters of the WDs to estimate their ages, and thereby provide a link
between these old objects and their low metallicity red subdwarf
counterparts.  Finally, we discuss the results in the context of
Galactic stellar populations.

\section{Observations}
\subsection{Photometry Observations and Reduction}

Initial $V_{J}R_{KC}I_{KC}$ (hereafter, subscripts not shown)
photometry observations for LHS 193AB and LHS 300AB were reported in
\citet{Jao2005}.  Data were taken at the CTIO 0.9m telescope, using a
1024$\times$1024 Tek CCD with image scale 0$\farcs$401 pixel$^{-1}$
during CTIOPI.  An additional $VRI$ observation set was recently
acquired for LHS 300AB, and improved deconvolution techniques allow us
to improve the photometry values for components in both systems.  The
aperture sizes applied to LHS 193B (aperture 4\arcsec, separation
12\farcs6) and LHS 300B (aperture 2\arcsec, separation 4\farcs3) in
\citet{Jao2005} were large enough to include contaminating flux from
the primary stars' wings.  A 1\arcsec~aperture size is now used for
the WD components in both systems, and the results are given Table 2.
These improved results indicate that the two white dwarfs are not
quite as red as reported in \citet{Jao2005}, although they are still
among the reddest WDs known, as can be seen in Figure 1.  In order to
test the reliability of the small aperture photometry, we compared the
results with those obtained by performing PSF fitting and removal of
the bright primary component of each system and then recalculating the
photometry.  Results from both methods were entirely consistent within
the observational errors.  The final signal-to-noise errors in $VRI$
for the red dwarf components are 0.001 mag, whereas for the WDs the
errors are 0.04 mag and the total errors are (including Signal to
noise, systematics and night-to-night variabily) 0.03 for the SDs and
0.05 for the WDs.

Infrared photometry of the LHS 193 and LHS 300 systems in the $JH$
bands (standard CIT/CTIO filters) have been obtained using the ANDICAM
detector at the CTIO 1.3m telescope via SMARTS time.  The IR Array for
ANDICAM is a Rockwell 1024$\times$1024 HgCdTe Hawaii Array with
18-micron pixels, providing an image scale of 0$\farcs$137
pixel$^{-1}$ and a field of view of 2$\farcm$4$\times$2$\farcm$4.
Observations made before February 2005 have a missing quadrant in the
upper right corner due to a chip malfunction that has since been
fixed.

The observations were carried out with the intent of doing
differential photometry using reference stars with well determined
2MASS magnitudes.  To optimize image quality we used a dithering
pattern with offsets of 15\arcsec. Because of the limited field of
view, there were only 1 and 3 suitable reference stars for LHS 193B
and LHS 300B, respectively, but all four reference stars have 2MASS
$JHK_{s}$ magnitude errors of less than 0.04. These reference stars
were also checked for photometric stability by using already existing
frames from the long term CTIOPI parallax project -- they showed no
signs of variability larger than 0.02 mag in the optical (and
presumably even less in the infrared).  The 2MASS photometry has been
converted to the CIT system by using the second generation relations
from \citet{Carpenter2003}\footnote{The first generation
transformations are from \citet{Carpenter2001}, based on the 2MASS
second incremental data.  After the 2MASS all sky database was
released, the transformations have been recalibrated and released on
Carpenter's website at
http://www.astro.caltech.edu/\~{}jmc/2mass/v3/transformations. }.

The images were corrected for bias and flat-fielded using standard
IRAF procedures. The reduced science frames were used to obtain
aperture photometry in Interactive Data Language (IDL). The reason for
adopting IDL in this step was the necessity to remove precisely the
contribution of the primary components in the LHS 193 and LHS 300
systems by fitting the non-uniform background and subtracting it. Even
in the case LHS 193, where the separation is 12$\farcs$6, the
contribution of the primary component's wings is substantial at the
location of the secondary, especially given the great magnitude
difference of these two components ($\sim$7 magnitudes at $J$). After
subtraction, the aperture photometry of the faint secondary components
and reference stars was completed using a series of apertures with
radii of 3 to 9 pixels in steps of 0.5 pixels. The final adopted
magnitude was the average of all these values with the uncertainty
given by the standard deviation of the mean (0.08 mag for $J$ and 0.1
mag for $H$).

\subsection{Spectroscopy Observations and Reduction}

Spectroscopic observations of LHS 193AB and LHS 300AB were made with
the 1.5-m and 4-m telescopes at CTIO, respectively.  For the
observations on the 1.5-m, the R-C spectrograph with a Loral
1200$\times$800 CCD camera was used with the \#32 grating (in first
order) at tilt 15.1$^\circ$.  The order-blocking filter OG570 was
used, resulting in spectra covering 6000\AA~to 9500\AA~with a
resolution of 8.6\AA.  At the 4.0-m, the R-C spectrograph with a Loral
3K$\times$1K CCD was used with the \#181 grating (in first order) at
tilt 58.8$^\circ$.  The order-blocking filter OG515 was used,
resulting in spectra covering 5500\AA~to 10000\AA~with a resolution of
6\AA. The fringing effect on the 4-m spectra at wavelengths longer
than $\sim$7000\AA~has been removed by customized IDL routines.
Reductions were carried out in the standard way using IRAF reduction
packages.  Wavelength and flux calibrations were done using {\it
twodspec.apextract} within IRAF.  Results are shown in Figure
\ref{spec_comp}.

Because of the wide separation of the LHS 193AB system, the spectra
were observed separately for each component.  However, the spectrum
labelled LHS 300A in Figure \ref{spec_comp} is actually a combined
spectrum.  The contamination is negligible given the large magnitude
difference between the two components ($\Delta V =$ 4.6, $\Delta R = $
4.9, $\Delta I = $ 5.0).  A spectrum of LHS 300B only was obtained by
carefully aligning the slit, although some contamination from the
primary is still present.

\section{Spectra of the Red Subdwarfs and White Dwarfs}

\begin{figure}[ht]
\begin{center}
\includegraphics[angle=90,scale=0.7]{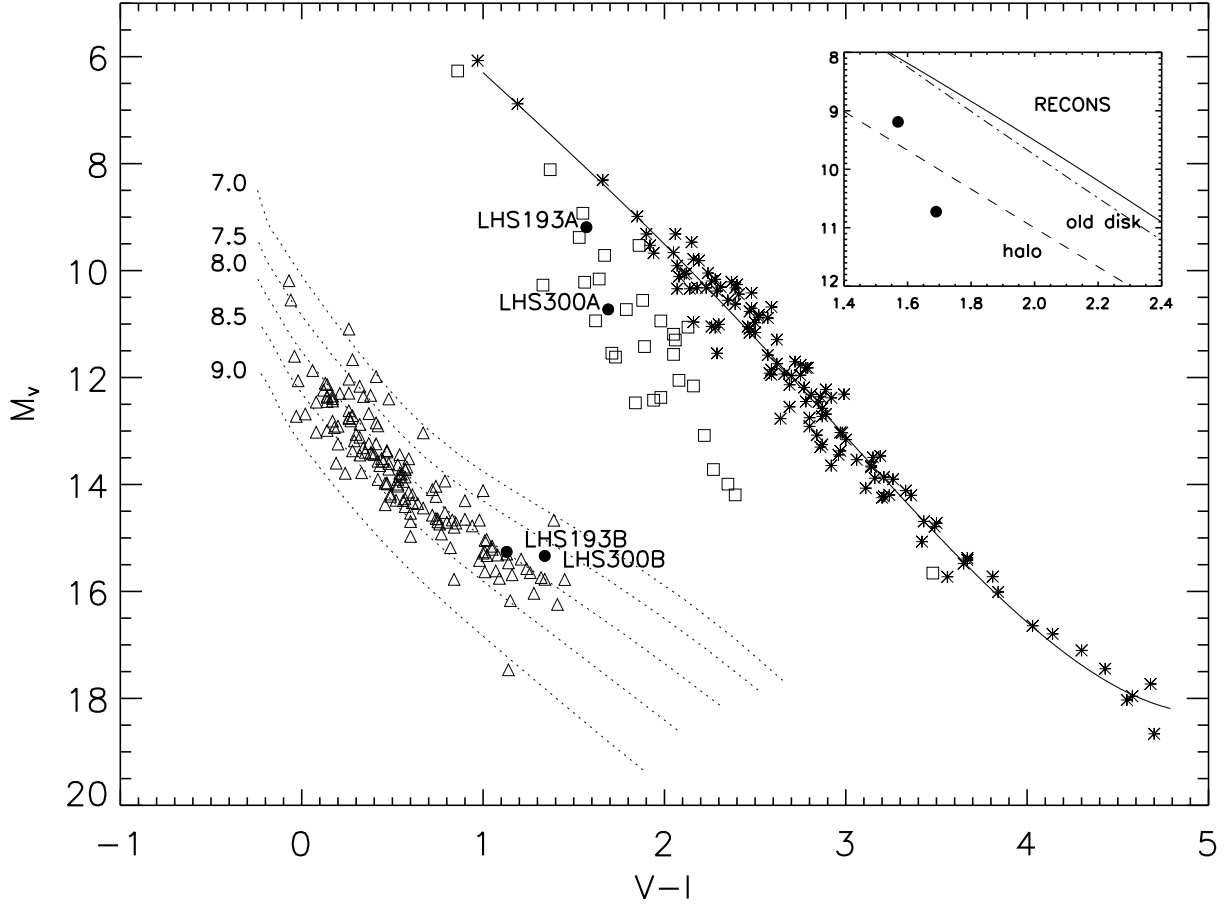}
\end{center}
\caption{Components of the LHS 193 and LHS 300 systems are labeled
with solid circles. The open squares represent subdwarfs with $\mu>$
1\farcs0 yr$^{-1}$ from \citet{Gizis1997}. Asterisks are RECONS stars
(stars within 10pc) used to outline the main sequence, and the solid
line indicates the best fitted line for these stars. Triangles are
white dwarfs from Bergeron et al (2001). The dotted line shows
different helium-rich cooling curves with $\log (g)$ labeled. The
inset is a zoom-in for both primaries. The solid line is the same as
defined above. The dash-dot and dash lines are best fits for old disk
and halo stars, respectively, from \citet{Leggett1992}.\label{hrdiag}}
\end{figure}

\begin{figure}[ht]
\begin{center}
\includegraphics[scale=0.95]{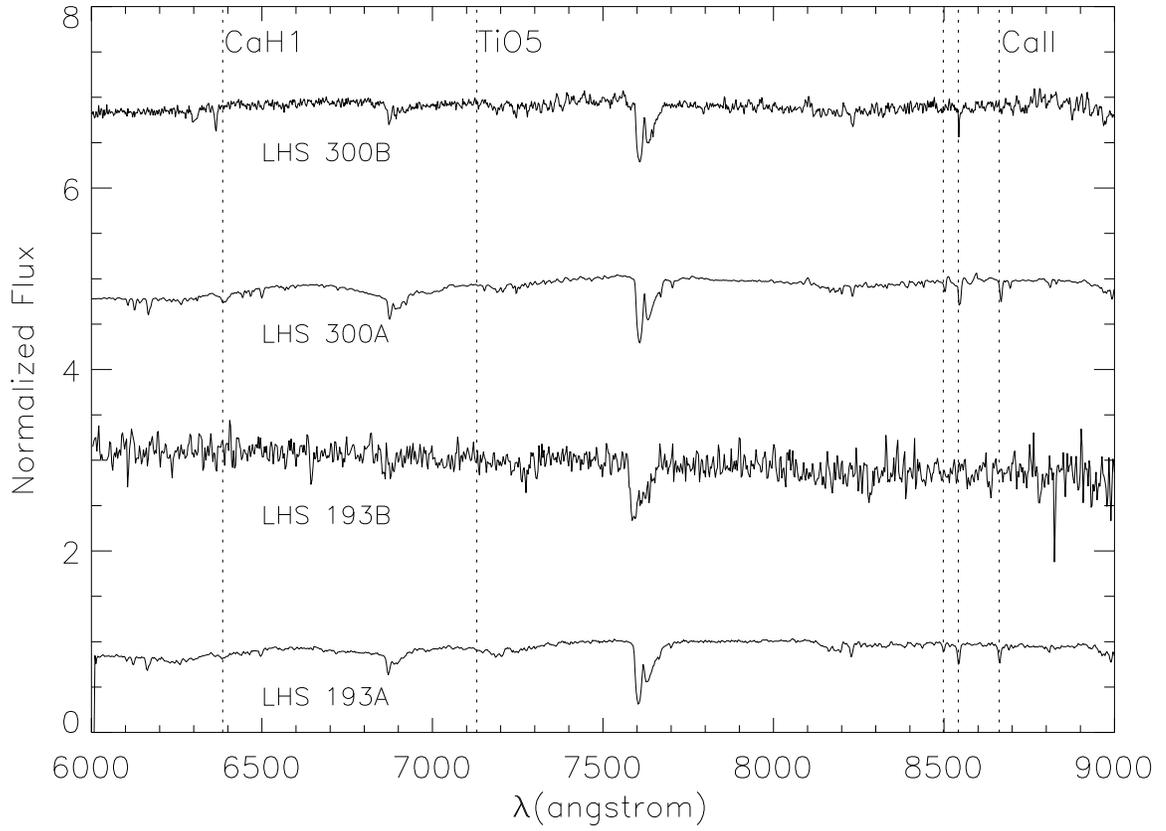}
\end{center}
\caption{ Spectra obtained for LHS 193AB and LHS 300AB normalized at
7500 A (offset for clarity).  These are combination of different
spectra taken with different instruments, see text for details. The
important absorption lines of CaII, as well as the center of the CaH1
and TiO5 bands are indicated with dotted lines. \label{spec_comp}}
\end{figure}

\begin{figure}[ht]
\begin{center}
\includegraphics[scale=0.7, angle=90]{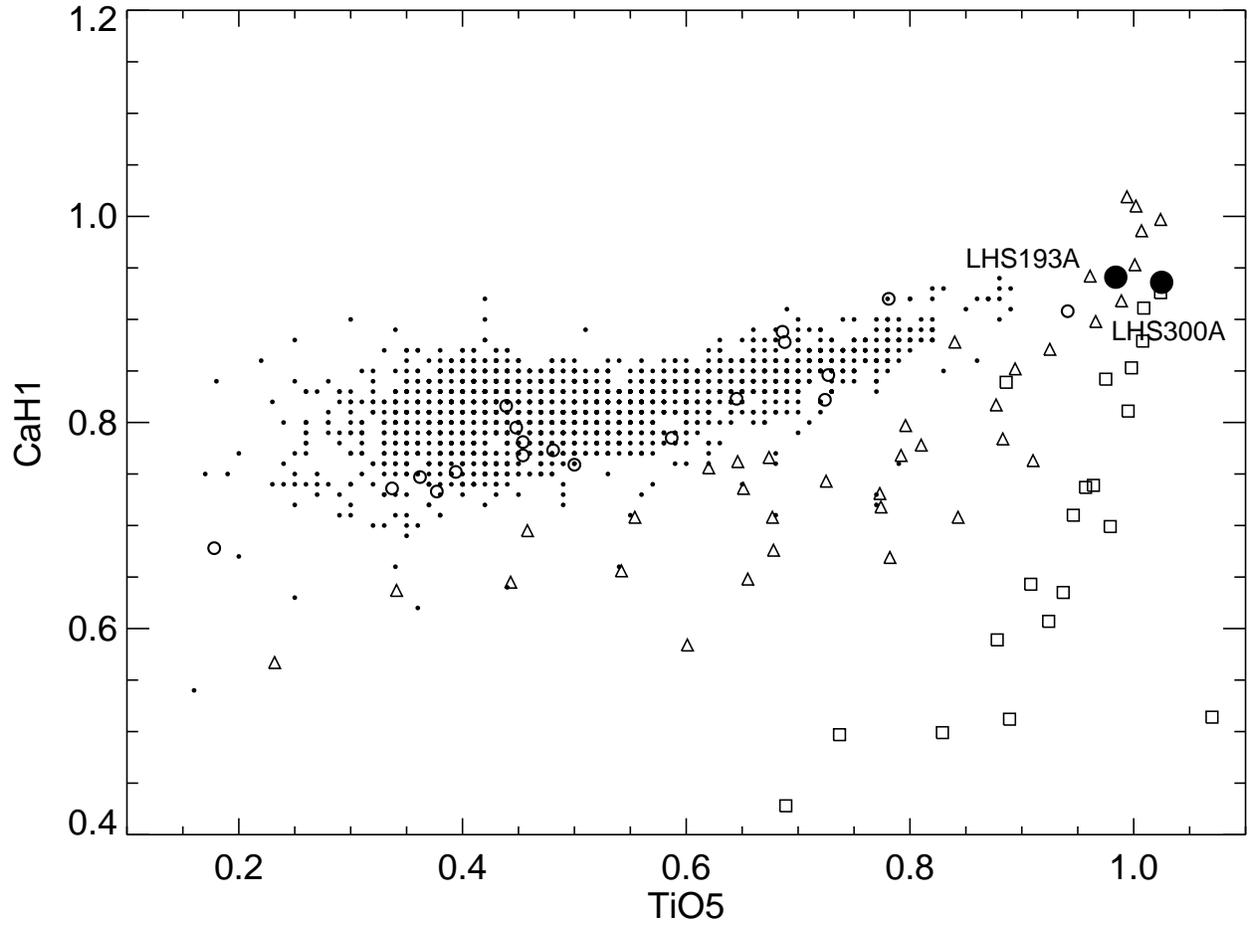}
\caption{Stars classified as main sequence stars from
\citet{Gizis1997} appear as open circles, subdwarfs as open triangles
and extreme-subdwarfs as open boxes.  Small dots represent main
sequence stars from \citet{Hawley1996} using the same indices.  Two
filled circles represent our two science stars, LHS 193A and LHS
300A.  The CaH1 band is from 6380\AA~to 6390\AA~and the TiO5 band is
from 7126\AA~to 7135\AA.\label{fig:CaH.TiO}}
\end{center}
\end{figure}

A sample of cool subdwarfs was previously studied by \citet{Gizis1997},
who showed that the subdwarf absorption bands of CaHn (n$=$1-3) and TiO
are the best features to separate subdwarfs from main sequence stars
between 6000\AA~and 9000\AA~(our spectral coverage).  Here we apply the
same spectroscopic index method for both primary stars, LHS 193A and LHS
300A.  Results are presented in Figure~\ref{fig:CaH.TiO}.  The
spectroscopic indices show that both stars fall in the early M type
subdwarf region.  When combined with the HR diagram results shown in
Figure~\ref{hrdiag}, the two pieces of evidence confirm, unequivocally,
that LHS 193A and LHS 300A are subdwarfs.

Spectra for each of the white dwarf secondaries are presented in
Figure 2.  Both spectra are noisy because these objects are faint, but
LHS 193B shows no apparent features except for the telluric lines seen
in all the spectra in Figure 2. LHS 300B does show some continuum
structure, especially in the red region of the spectrum due to
contamination from the bright primary. One prominent feature seen
close to $\lambda~8542$ does not have the same width as the
corresponding one in the primary component and no trace of the other
two lines of the CaII triplet, to which it should belong, are present,
indicating that it is probably due to residual cosmic rays. These
objects would be traditionally classified as featureless DC white
dwarfs.  Ideally, higher S/N spectra are desirable to deconvolve any
trace hydrogen absorption from the noise should it exist (H$\alpha$ is
within this spectral coverage at 6563 A).  We tentatively conclude
that neither WD spectrum has hydrogen absorption features, which
justifies the use of helium-rich models to characterize these objects,
as discussed in the next section.

\section{Ages of the White Dwarf Components}

One of the most interesting aspects of WDs in general is that they
provide us with a means of determining the age of the system to which
they belong.  This is possible because the physics of WD cooling is
relatively well-understood and unhindred by assumptions about nuclear
reaction rates, among others, allowing for reliable models to be
constructed.  Many models have been published in the literature for
the cooling of WDs over the past several years (see for example
\citealt{W90},\citealt{W95},\citealt{BA98},\citealt{H99} and
\citealt{MKWW99} among others). Our goal in this section is to
estimate the {\it total} ages of the LHS 193 and LHS 300 systems by
summing each WD's cooling age and its main sequence lifetime.  The set
of models described in \cite{BWB95} is used here to determine the
cooling ages, as well as other stellar parameters.  The choice of this
model in particular is due to the fact that the author provides
complete grids for various photometric filters and physical quantities
of the WDs, making it possible to use interpolation procedures to get
physical parameters from observed photometrical values. With these
results, we utilize the initial to final mass relations of \cite{W92}
and \cite{IL89} to estimate the progenitor masses and consequent main
sequence lifetimes. A modern revision of such relations is presented
by \cite{W00} where they suggest a new form for this
function. However, this revised relation is not significantly
different from the one of \cite{IL89}, which we adopt here for
consistency with the discussion of \cite{W92}.

\subsection{White Dwarf Physical Parameters and Cooling Ages}

The models of \cite{BWB95} provide us with bolometric corrections and
color indices on various photometric systems calculated for an
extensive grid of hydrogen- (DA)and helium-rich (non-DA) white dwarf
model atmospheres.  Absolute visual magnitudes, masses, and ages are
obtained for each model from detailed evolutionary cooling sequences.
For each set of filters, grids are calculated for DA and non-DA
atmosphere models.  The DA model grid covers a range from $T_{eff} =
1500 K$ to $100000 K$ and from $log (g)$ = 7.0 to 9.0, while the
non-DA model grid covers a range from $T_{eff} = 3500 K$ to $30000 K$
and from $log (g)$ = 7.0 to 9.0.  The grids are available for public
use at: {\it
http://www.astro.umontreal.ca/\~{}bergeron/CoolingModels/}.

One important step in this procedure is to decide which set of models to
use for the interpolation required to match observed parameters for each
WD.  Photometry alone is not enough to decide between DA or non-DA model
atmospheres and spectroscopy should always be available to pinpoint the
type of WD being studied.  This is very important, especially for age
determinations, because cooling times are strongly dependent on the type
of model considered, i.e.~non-DA models give different cooling ages for
a given position in the ($V-I$)$\times$$M_{V}$ plane than the DA models.
In the case of LHS 193B and LHS 300B, both spectra show no evidence of
hydrogen lines (as discussed in section 3), indicating very low abundance
of the element in the atmosphere.  Thus, they were both studied with He
model atmospheres.

We estimate the physical characteristics and ages of the WD components
using the $VRIJH$ photometry for LHS 193B and LHS 300B.  To map the
WDs into the model grids, we interpolated values in the ($V-R$,
$M_{V}$), ($V-I$, $M_{V}$), ($V-J$, $M_{V}$), ($V-H$, $M_{V}$) planes
for each object to find their surface gravities, effective
temperatures, masses, and cooling ages.  The independent
determinations from these color combinations are then averaged to
obtain a final value for each parameter.  The routines take the
regularly-gridded set of models and use a smooth quintic surface for
the interpolation. The result is s a two-dimensional floating-point
array containing the interpolated surface, sampled at the grid
points. The color amd absolute magnitude combinations discussed above
are then interpolated on this surface using cubic convolution
interpolation. For more detail on the interpolation routines see IDL
manual and documentation. Results are given in Table 3.

\begin{figure}[ht]
\begin{center}
\includegraphics[width=\columnwidth]{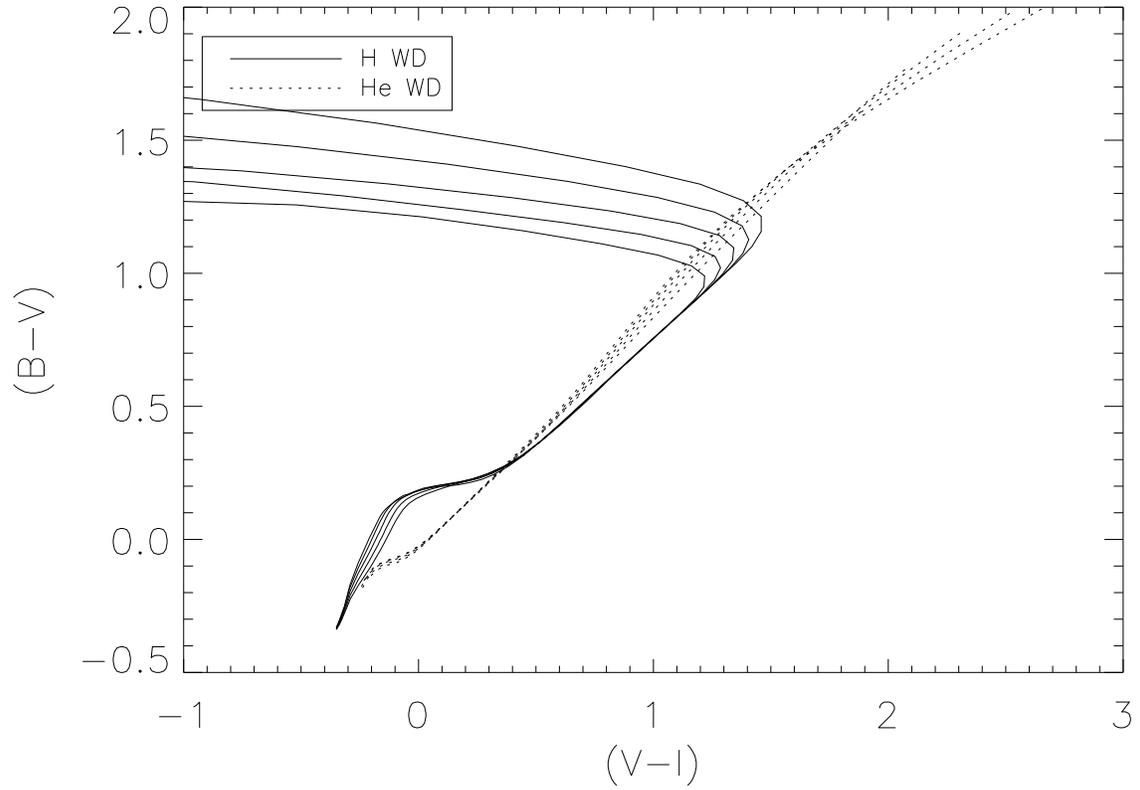}
\end{center}
\caption{ Model curves for DA (H models) and non-DA (He models) WDs
for different $log (g)$ values from 7.0 (top curve for DA models) to
9.0 (bottom curve for DA model) showing the degeneracy portions of
this particular color-color relation (see text for details on the
models used). The curves for the non-DA models cover the same $log
(g)$ range as the H models.
\label{mod_degen}}
\end{figure}

The reason for utilizing the values for ($V-R$), ($V-I$), ($V-J$),
($V-H$) and $M_{V}$ is simply due to the nature of the model grids
available and their properties.  Other relations can be constructed
but are of limited use because of overlapping curves for different
surface gravity values, thereby introducing degeneracies.  As an
example, one could use the ($V-I$, $B-V$) relation if these
photometric values are available.  However, if we plot the model grid
for these parameters, we see that the curves for different surface
gravity values overlap for some sections of the diagram (as shown in
Figure~\ref{mod_degen}).  Consequently, this particular color-color
diagram is degenerate and would not provide meaningful results.

To determine the reliability of our method we compared our results
with those of a ``control'' WD, GJ 283A with type DQ, which is the
primary star of a binary system with an M6.0V main sequence red dwarf
companion.  This provides two opportunities for comparison --- the WD
has also been evaluated by \cite{BLR01} so we can check the
consistency of results, and the main sequence red dwarf is likely much
younger than the red subdwarfs, LHS 193A and LHS 300A, so we should
determine a younger cooling age for GJ 283A.  Using our method and the
observed values from \cite{BLR01} for GJ 283A, we find differences of
1\% in surface gravity, 4\% in temperature, 14\% in mass and 10\% in
cooling age.  These differences are likely due to the nature of the
interpolation process and on model grid resolution; we conclude that
our procedure is sound.  Derived cooling ages and other physical
parameters for all three white dwarfs are listed in Table 3.  We
confirm that the GJ 283 system is significantly younger than the LHS
193 and LHS 300 systems.

\subsection{White Dwarf Progenitor Masses and Main Sequence Lifetimes}

Using relations presented in \cite{W92}, we can estimate the
progenitor masses of the WDs and their main sequence lifetimes via the
following equations:

\begin{equation}
M_{WD}=A\times e^{(B\times (M/M_{\odot}))}
\end{equation}
 
\begin{equation}
t_{MS}=10 \times (M/M_{\odot})^{-2.5} Gyr
\end{equation}

\noindent where $M_{WD}$ is the mass of the white dwarf, A and B are
constants (see description below), $M$ is the mass of the progenitor
(both masses in solar units), and $t_{MS}$ is the main sequence
lifetime in $Gyr$.

The constants in Equation 1 need to be defined so that a value of $M$
can be determined.  In this work we adopted three different
prescriptions for these constants, two from \cite{W92} and one is from
\cite{IL89}.  The two prescriptions from \cite{W92} are the ones
referred to in that work as Model D ($A=0.40$ and $B=0.125$) and Model
E ($A=0.35$ and $B=0.140$).  Model D in particular is what the author
defines as the ``Best Guess Model''.  The third prescription uses the
Initial--Final mass relation defined by \cite{IL89}.  In this work we
refer to these different prescriptions as Wood D, Wood E and Iben 89,
respectively.  The choice of these particular models is based on the
theoretical mass functions obtained from them.  As mentioned in
\cite{W92}, the best fit to the observed mass function is achieved
using Wood D.  Wood E and Iben 89 are included as independent checks
on the progenitor masses and main sequence lifetimes.  Results for all
three methods are given in Table 3.

\subsection{White Dwarf Total Ages}

The total ages of LHS 193B and LHS 300B are simply the sums of their
main sequence lifetimes and their cooling ages, i.e.~$t_{cool}$ and
$t_{MS}$ given in Table 3. Given the various uncertainties in the
methods used to determine both portions of the WDs' $total$ ages, we
conclude that both systems have ages of 6-9 Gyr.  By proxy, we
conclude that the low metallicity subdwarf components have the same
ages as their WD partners, thereby providing a link between WD ages
and red dwarf metalicities.

\section{The Rarity of Cool Subdwarf-White Dwarf Binaries}

Given the utility of these WD-subdwarf pairs in investigating the link
between age and metallicity, we have carried out searches for these
systems in Sloan Digital Sky Survey (SDSS), the L{\'e}pine Shara Proper
Motion (LSPM, see \citet{Lepine2005} for details) northern hemisphere
catalog and the wide commom proper motion pairs reported by
\citet{Chaname2004}. In the SDSS data, \citet{West2004} have detected 60
subdwarfs from the analysis of nearly 8000 late type dwarfs.  We compare
the positions of these spectroscopically confirmed subdwarfs with the
SDSS WD catalog of \citet{Kle2004} to reveal WD-subdwarf pairs.  The WD
catalog has coordinates (equinox 2000.0, epoch approximately 2000.0) and
many proper motions for the objects, but the subdwarf catalog only has
coordinates (equinox 2000.0).  In order to find potential close binaries
in the two catalogs, a coordinate comparison was carried out using a
generous 5\arcmin~search radius around each entry of the WD catalog.  No
subdwarfs were found in \citet{West2004} that were within 5\arcmin~of
any SDSS WD catalog entries.

\begin{figure}[ht]
\begin{center}
\includegraphics[scale=0.7, angle=90]{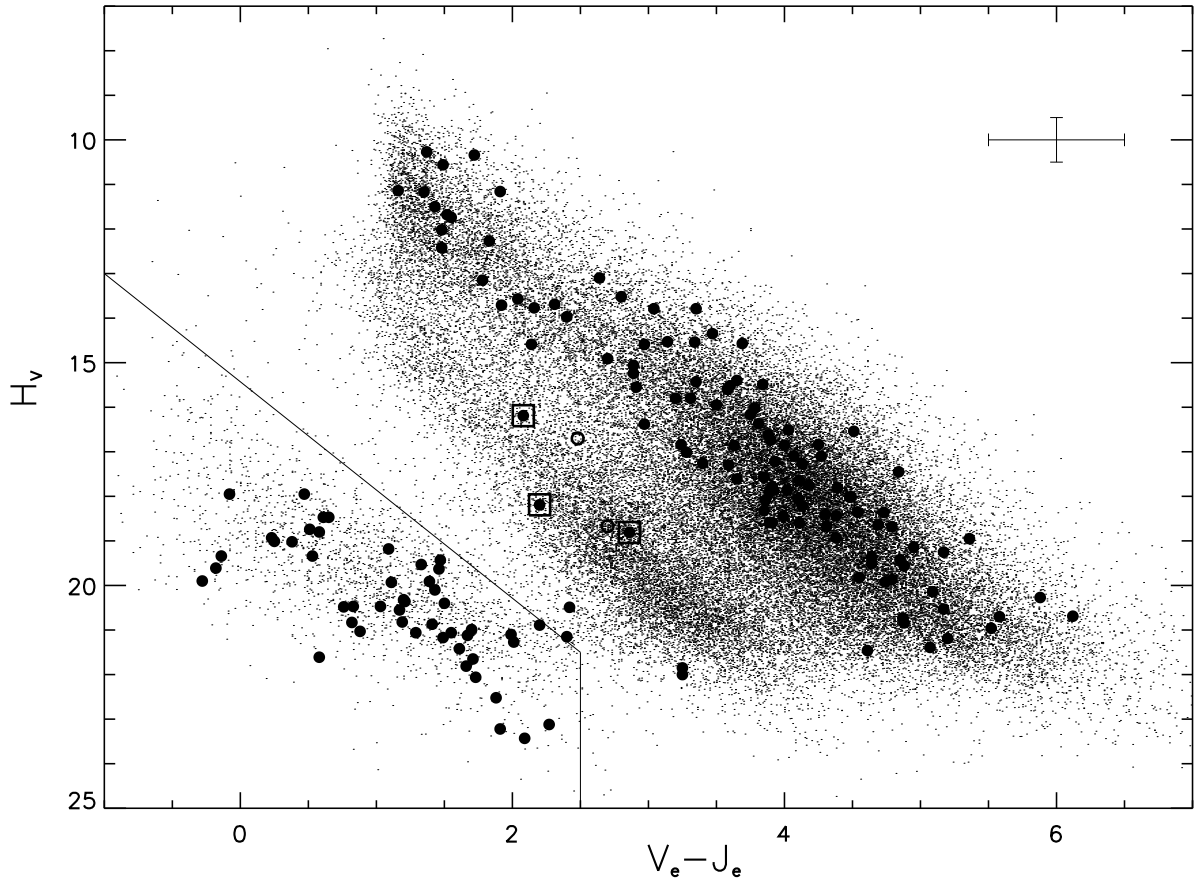} 
\caption{Reduced proper motion diagram for LSPM catalog.  Three
different luminosity populations can easily been seen.  Stars to the
lower left of the thin solid line are selected as ``WD'' candidates.
Stars that are companions (less than 1\arcmin~separation) to these
``WD'' are indicated as filled circles.  Three boxed stars have
observed $J$ from 2MASS.  The two open circles indicate LHS 193A (top)
and LHS 300A (bottom) using $V$ and $J$ magnitudes reported
here. The error bars are shown in the upper right side.\label{lspm.RPM}}
\end{center}
\end{figure}

\citet{Lepine2005} have released the most comprehensive catalog of
high proper motion stars in the northern sky, which contains nearly
62,000 stars with proper motions greater than 0\farcs15 yr$^{-1}$.
Populations of main sequence stars, subdwarfs, and white dwarfs found
in the catalog can be separated effectively (although not perfectly)
using proper motions and magnitudes to calculate each object's reduced
proper motion, H, as introduced by \citet{Luyten1939} and defined as
follows,

\begin{equation}
H=m+5\log\mu+5,
\end{equation}

\noindent where $\mu$ is the proper motion in seconds of arc per year
and $m$ is the apparent magnitude, $V$.  The simple assumption is that
proper motion is correlated with parallax, so that samples without
parallax measurements can be investigated.  \citet{Gizis1999},
\citet{Digby2003} and \citet{Salim2003} have previously used this
technique to separate subdwarfs from main sequence stars.  Our search
of the LSPM catalog utilizes the reduced proper motion diagram in
Figure~\ref{lspm.RPM}, which plots $H_V$ vs.~$V-J$ color.  Because
many faint objects in the LSPM catalog lack $V$ band photometry,
\citet{Lepine2005} estimated the optical $V$ band magnitudes from
photographic plate $B_{J}$ and $R_{F}$ magnitudes.  The vast majority
of LSPM have reliable near infrared photometry from 2MASS.  For those
that do not, \citet{Lepine2005} estimate $J$ from plate $I_{N}$
magnitudes.

We have defined rough boundaries, plotted in Figure~\ref{lspm.RPM},
to indicate the WD region.  All possible WD candidates within this
region are compared with all other stars in the LSPM catalog to reveal
companions, with the particular goal of finding subdwarf companions.
The LSPM catalog has a coordinate accuracy of about $\sim$90 mas and
proper motion accuracy of $\sim$10 mas yr$^{-1}$.

Two common proper motion binary searches were done, using the
following criteria:
\begin{enumerate}

\item  Separation of WD and candidate companion less than 1\arcmin,
proper motion difference less than 0\farcs1 yr$^{-1}$, and position
angle of the proper motion difference less than 5 degrees;

\item Separation of WD and candidate companion between
1\arcmin~and 5\arcmin, proper motion difference less than 0\farcs01
yr$^{-1}$, and position angle of the proper motion difference less
than 1 degree.
\end{enumerate}

The results from the first search method are plotted in
Figure~\ref{lspm.RPM}.  Companions of the WD candidates are shown as
filled circles.  There are 171 total binary system candidates, 8 of
which are not shown on this figure because they lack $V-J$ colors.  Of
the 163 systems with colors, 108 companions are in the main sequence
region, 48 companions are in the WD region (probable double degenerate
binaries), and only seven stars are in the subdwarf region.  These
seven systems are the WD-subdwarf systems of interest here, and are
listed in Table 1.  Of these seven, four have estimated $V$ and/or $J$
magnitudes and are on the edges of the subdwarf region, whereas three
have good photometry from 2MASS.  We will carry out spectroscopic
follow up observations to determine their luminosity classes. The
second search revealed 24 candidate common proper motion binaries.
Twenty of the companions are in the main sequence region, four are in
the WD region, and none are in the subdwarf region.

The third database from \citet{Chaname2004} Table 1 has been searched
for the sdM-WD binaries. Although there are 1147 pairs of wide commom
proper motion binaries, only 58 systems been classified as ``at least
one component is a white dwarf'' by them based on the reduced proper
motion diagram. One system 880-460/880-461 (NLTT catalogue identifiers)
is shown to be sdM-WD binary on the reduced proper motion
diagram. However, after carefully checking the photographic plates
having about 14 years time span, both stars do not appear to move at
0$\farcs$25 yr$^{-1}$ and both ``components'' do not have common proper
motions as reported in the NLTT catalogue. Therefore, this system is not
convincing and no sdM-WD has been found from Table 1 in
\citet{Chaname2004}.

We conclude that WD-subdwarf pairs are quite rare.  The three searches
revealed only seven candidate systems, and only three of those have
reliable photometry.  Accurate $V$ and $J$ photometry for more of the
stars in LSPM might reveal additional candidate systems.

\section{Discussion}

The rarity of subdwarfs in the solar neighborhood increases our
difficulty in understanding their physical parameters, including their
sizes, masses, and even ages.  The discovery of the two rare
WD-subdwarf systems reported here provides a crucial step towards a
better understanding of the star formation history of our Galaxy
because these systems allow us to link the ages of the WDs to the
metallicities of the subdwarfs.  Figure 3 indicates that the
metallicities of both LHS 193A and LHS 300A are [m/H] $\sim$ $-$0.5
(details of the methodology are the subject of a future paper).
Unfortunately, these systems are rare, so it will take considerable
effort to build up a statistical database.  Nevertheless, a careful
reconnaissance may reveal enough systems to produce a map of age
vs. metallicity that will provide valuable insight into the Galaxy's
past.

The HR diagram and spectroscopic analysis both indicate that the red
components in the WD-subdwarf systems are true subdwarfs. From our
astrometry results of the subdwarf components (\citealt{Jao2005}), we
also know that the tangential velocities are indicative of an old
population --- 147 km/sec for LHS 193A and 183 km/sec for LHS
300AB. Comparison of our results to the kinematic study of low mass
stars of \citet{Leggett1992} (shown in the inset of
Figure~\ref{hrdiag}), both primaries appear to have halo-like
kinematics. The WD {\it total} ages of 6-9 Gyr derived here indicate
that the systems are not members of the young disk or extremely old
halo, but they are probably members of the old disk (or
young halo) population.

To estimate the $total$ ages discussed above we calculated values of
progenitor mass and main sequence lifetimes for two prescriptions of
\cite{W92} as well as the relation proposed by \cite{IL89}.  When
comparing the progenitor masses determined for each star we see that
there is a considerable difference when adopting one or another
prescription for the initial to final mass relation.  However, the
final ages obtained are within the approximate error of 1 Gyr
suggested by \cite{W92}.  This is a clear indication that although
these prescriptions should be used with caution, reliable total age
estimates are possible because the cooling time is the dominant
factor.  Figure 6 shows the three systems studied here in the
$M/M_{\odot} \times T_{eff}$ diagram, with a background of cooling
curve isochrones (solid lines) and corresponding $total$ age curves
(dotted lines) for the ``Best Guess Model'' of \cite{W92}.  Notice
that LHS 300B is located at the convergence region of the $total$ age
curves and therefore it could be even older than the 6-9 Gyr estimate,
given the uncertainty in its mass.

Uncertainties in WD cooling times and temperatures may cause the ages
reported here to be in error.  \cite{FBB01} discuss in detail many
current issues in WD cooling time calculations.  Of particular
interest to our work is the variation of cooling curves due to
different adopted abundances for WD cores.  The authors point out that
discrepancies as large as 3.6 Gyr can occur for a given luminosity of
$10^{-6}~L_{\odot}$ by adopting either a pure O core or a pure C
core. However, the cooling ages determined in this work are in the
range of 1--7 Gyr, where this discrepancy should still be within our
method uncertainty of ~1 Gyr (see Figure 7 in \cite{FBB01}).  As an
example of the kind of changes that may occur in WD models, Dufour et
al (2005) showed that the temperatures of DQ WDs may be significantly
lower than previous values, when model atmospheres including metals
and molecules are considered.  This in turn has a direct implication
on the masses determined for these objects.  They find that the mean
masses of the DQ WDs are reduced from 0.72 to 0.61 $M/M_{\odot}$, in
better agreement with the non-DA WD mean mass.  This effect may also
be important for non-DA WD stars in general, as in our case,
indicating that the ages estimated here should be taken as an upper
limit.

The only other WD-subdwarf system with an estimated {\it total} age
for the WD is LP164$-$51/52, for which \cite{Silvestri2005} identify
the system as a halo candidate with an age estimate of 9.6 Gyr
(Silvestri, private communication).  This age is based on an assumed
WD mass of 0.61 $M_{\odot}$, and an interpolation of the same model
grids used here, although for the ($V-I$ vs Age) and/or ($B-V$ vs Age)
diagrams.  However, this procedure could lead to erroneous age
estimates if the mass value is incorrect by even 0.1 $M_{\odot}$, as
can be seen, for example, in our Figure 6 for a given temperature.
Ideally, a trigonometric parallax will be determined for LP164$-$51/52
and the characterization of the system can be completed.

\section{Conclusions}

In this work we present accurate $VRIJH$ and spectroscopy from
6000\AA~to 9000\AA~for two new WD-subdwarf systems, LHS 193AB and LHS
300AB.  Astrometry and photometry clearly identify each system to be
comprised of a WD and a red subdwarf.  The featureless spectra
indicate that both WDs are type DC. In combination with our
astrometry, photometry, and spectroscopy, we use literature model
grids for helium-rich atmospheres to determine the surface gravities,
temperatures, masses, and cooling ages of the WDs.

With the cooling ages and masses of the WDs, we estimate progenitor
masses and main sequence lifetimes for the objects, and add the
cooling ages to derive $total$ ages of 6-9 Gyr for both systems.  To
determine the progenitor masses, we use three different prescriptions
to avoid biases toward any particular model, and confirm the
approximate $\sim$1 Gyr uncertainty in the ages suggested by
\cite{W92}.  Although some work has been done on the subject of
mapping the connection from progenitor mass to remnant mass
\citep{W87,WK83} there is still not a reasonable final solution to
this question and the values presented here may change with different
models.  We do not consider this to be a significant problem because
the cooling age comprises most of the $total$ age.

These are the first two WD-subdwarf system for which precise
photometry, spectroscopy, and most importantly, parallaxes, are
available.  This unique combination allows us to link the ages of the
WDs to the low metallicities of specific red subdwarfs for the first
time.  We find that these two systems are likely members of the thick
disk population of the Galaxy, which is supported by the systems'
large tangential velocities.  Even when errors are considered, the
systems are not likely to be members of the halo because the $total$
ages are below the canonical age of 12 to 14 Gyr usually adopted for
halo type objects (see \cite{GWK89} and references therein for
details).  Accurate Radial velocity measurements, when combined with
the accurate astrometry reported here, would also provide useful
insights into the kinematic population of the two systems. Nonetheless,
we find the compelling result that there exist ancient stars in the
solar neighborhood of low metallicity that are fossils of the early
star formation epochs of our Galaxy.

\begin{acknowledgements}

We appreciate the assistance of the members of the SMARTS Consortium,
Rebeccah Winnick in particular, and staff at CTIO who made many of the
observations reported here possible.  The RECONS team at Georgia State
University is supported by NASA's Space Interferometry Mission and
GSU.  This work has used data products from the Two Micron All Sky
Survey, which is a joint project of the University of Massachusetts
and the Infrared Processing and Analysis Center at California
Institute of Technology funded by NASA and NSF.

\end{acknowledgements}


\clearpage


\begin{table}
\centering
\caption{WD-subdwarf binary candidates\label{candidate.tbl}}
\begin{tabular}{lccc}
\tableline
\tableline
 Object                                &     $VRI$ Photometry   &    Spectroscopy          &   Parallax        \\
\tableline
LHS193AB                               &     this work          &     this work            &    \citet{Jao2005}\\
LHS300AB                               &     this work          &     this work            &    \citet{Jao2005}\\
LHS2139/2140\tablenotemark{a}          &     no                 &    \citet{Gizis1997-2}     &    no	       \\
GJ781AB\tablenotemark{b}               &     \citet{Weis1996}   &    \citet{Gizis1998}     &    \citet{YPC}    \\
LP164$-$51/52\tablenotemark            &     no                 &    \citet{Silvestri2002} &    no	       \\
LSPMJ0008$+$1634/1635                  &     no                 &     no                   &    no 	       \\
LSPMJ0846$+$4925E/W                    &     no                 &     no                   &    no	       \\
LSPMJ0927$+$6210N/S                    &     no                 &     no                   &    no	       \\
LSPMJ1200$+$4105N/S                    &     no                 &     no                   &    no	       \\
LSPMJ1209$+$2448E/W                    &     no                 &     no                   &    no	       \\
LSPMJ1539$+$5402N/S                    &     no                 &     no                   &    no	       \\
LSPMJ1702$+$7158N/S                    &     no                 &     no                   &    no	       \\
LSPMJ2100$+$3426E/W                    &     no                 &     no                   &    no             \\
\hline
\end{tabular}
\tablenotetext{a}{ LHS 2140 has spectral features of an M-subdwarf, but
LHS 2139 spectrum appears noisy but featureless.}

\tablenotetext{b}{GJ 781AB is a spectroscopic binary and has combined
photometry reported in \citet{Weis1996}. GJ 781A, an M subdwarf, shows
absorption lines broader than regular M dwarfs.}

\end{table}

\clearpage


\begin{table}
\begin{center}
\caption{$VRIJH$ photometry for binary systems.\label{obs-res}}
\begin{tabular}{cccccccccc}
\hline  
{object}    &   {$V$}   &   {$R$}  &   {$I$}  & $\#$ obs & {$J$}    &   {$H$}  & $\#$ obs &   $\pi$(mas)   &  Ref                \\
\hline										      
\hline 										      
{LHS 193A}  &   {11.66} &  {10.85} & {10.09}  &  3       & {9.17}   &   {8.55} &  3       &  32.06$\pm$1.65&\citealt*{Jao2005}   \\
{LHS 193B}  &   {17.73} &  {17.18} & {16.60}  &  3       & {16.21}  &  {15.94} &  3       &    \nodata     &                     \\
{LHS 300A}  &   {13.18} &  {12.28} & {11.49}  &  2       & {10.48}  &  {10.00} &  2       &  32.30$\pm$1.85&\citealt*{Jao2005}   \\
{LHS 300B}  &   {17.79} &  {17.13} & {16.45}  &  2       & {15.94}  &  {15.89} &  2       &    \nodata     &                     \\
{GJ 283A}   &   {13.01} &  {12.86} & {12.71}  &\nodata   & {12.69}  &  {12.60} &\nodata   & 112.4$\pm$2.7  &\citealt*{YPC}       \\
{GJ 283B}   &   {16.54} &  {14.68} & {12.43}  &\nodata   & {10.14}  &  {9.61}  &\nodata   &    \nodata     &                     \\
\hline
\end{tabular}

\tablecomments{The $J$ and $H$ photometry have been converted from
the 2MASS to CIT system, except for LHS 193B and LHS 300B, which were
observed here and are already CIT. The optical photometry for GJ283 A and B is from
\citet{Bessel1990}.}

\end{center}
\end{table}

\clearpage


\begin{deluxetable}{lccccccccccccc}
\rotate
\tabletypesize{\scriptsize}
\tablecaption{Physical parameters and Ages for Studied Systems\tablenotemark{a}.\label{age-table}}
\tablehead{
\colhead{} &
\colhead{} &
\colhead{} &
\colhead{} &
\colhead{} &
\multicolumn{3}{c}{Wood D}  & 
\multicolumn{3}{c}{Wood E}  & 
\multicolumn{3}{c}{Iben 89}\\

\colhead{object}  &
\colhead{Log(g)}  & 
\colhead{Teff}    & 
\colhead{$M_{WD}$}& 
\colhead{${t_{cool}}$}   &   
\colhead{$M$ ($M_{\sun}$)}   & 
\colhead{${t_{MS}}$}   &   
\colhead{${t_{total}}$}         &
\colhead{$M$ ($M_{\sun}$)}   & 
\colhead{${t_{MS}}$}   &   
\colhead{${t_{total}}$}         &
\colhead{$M$ ($M_{\sun}$)}   & 
\colhead{${t_{MS}}$}   &   
\colhead{${t_{total}}$}
}

\startdata
{LHS 193B}                  &   {8.1}    &   {4934}  & {0.60}    &   {6.5}  &   {3.3}    &   {0.5}  &   {7.0}   & {3.9}    & {0.3}    &   {6.8}   &   {2.0}     &  {1.7}  &  {8.2}   \\
{LHS 300B}                  &   {7.8}    &   {4705}  & {0.48}    &   {5.0}  &   {1.4}    &   {4.2}  &   {9.1}   & {2.2}    & {1.4}    &   {6.3}   &   {$<$2.0}  &  {>1.8}  &  {6.7}   \\
{GJ 283A}                   &   {8.3}    &   {8457}  & {0.77}    &   {1.6}  &   {5.3}    &   {0.2}  &   {1.7}   & {5.8}    & {0.1}    &   {1.7}   &   {4.1}     &  {0.3}  &  {1.9}   \\
{GJ 283A\tablenotemark{b}}  &   {8.2}    &   {8010}  & {0.72}    &   {1.6}  &   {4.8}    &   {0.2}  &   {1.8}   & {5.2}    & {0.2}    &   {1.7}   &   {3.5}     &  {0.4}  &  {2.0}   \\
{GJ 283A\tablenotemark{c}}  &   {8.09}   &   {7710}  & {0.63}    &   {1.45} &   \nodata  &   \nodata&   \nodata &\nodata   & \nodata  &   \nodata &   \nodata   &  \nodata&  \nodata  \\
\enddata

\tablenotetext{a}{All ages and cooling times given in Gyr}
\tablenotetext{b}{these are values obtained with our interpolation procedure
using Bergeron (2001) observed values for GJ 283A.}
\tablenotetext{c}{Values obtained by \cite{BLR01}}

\end{deluxetable}

\clearpage


\clearpage


\clearpage


\clearpage


\clearpage


\clearpage


\begin{figure}[ht]
\begin{center}
\includegraphics[scale=0.8]{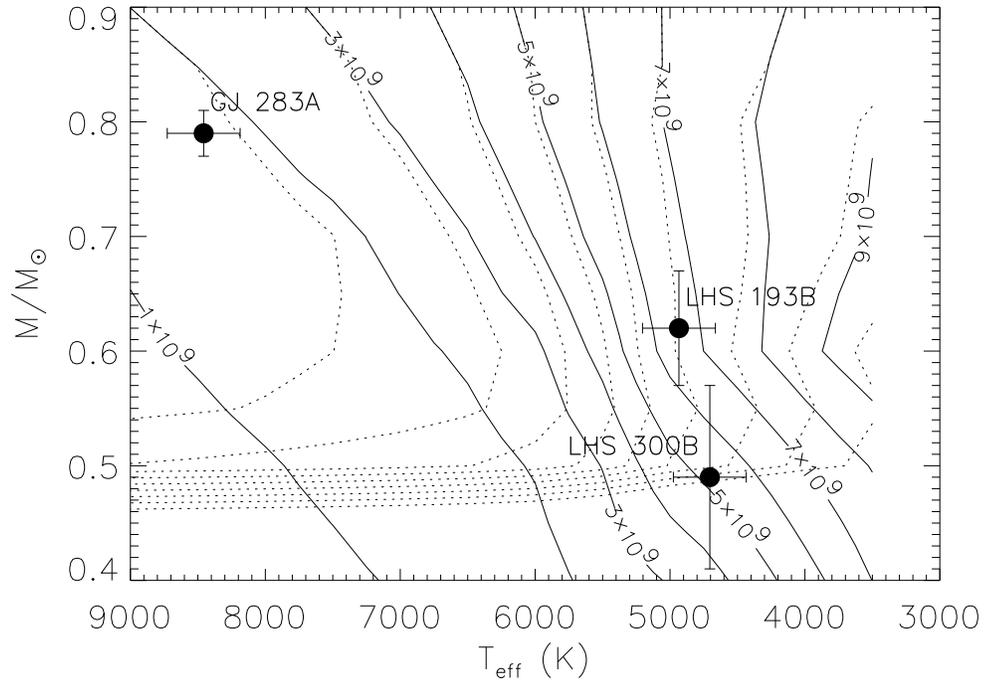}
\end{center}
\caption{ Position of the three systems studied in the $M/M_{\odot}
\times T_{eff}$ diagram.  Solid lines are the cooling age isochrones
and dotted lines are $total$ age isochrones (cooling age plus MS age)
with units given in years. The dotted lines are drawn for the same
isochrones values as the solid lines. \label{teff_mass_he}}
\end{figure}  


\begin{thebibliography}

\bibitem[Benvenuto \& Althaus(1998)]{BA98} Benvenuto, O.~G., 
\& Althaus, L.~G.\ 1998, \mnras, 293, 177

\bibitem[Bergeron et al.(1995)]{BWB95} Bergeron, P., Wesemael, F., \&
Beauchamp, A.\ 1995, \pasp, 107, 1047

\bibitem[Bergeron et al.(2001)]{BLR01} Bergeron, P., Leggett, S.~K.,
\& Ruiz, M.~T.\ 2001, \apjs, 133, 413

\bibitem[Bergeron et al.(2005)]{Bergeron2005} Bergeron, P., Ruiz,
M.~T., Hamuy, M., Leggett, S.~K., Currie, M.~J., Lajoie, C.-P., \&
Dufour, P.\ 2005, \apj, 625, 838

\bibitem[Bessel(1990)]{Bessel1990} Bessel, M.~S.\ 1990,
\aaps, 83, 357

\bibitem[Carney et al.(1994)]{Carney1994} Carney, B.~W., Latham, 
D.~W., Laird, J.~B., \& Aguilar, L.~A.\ 1994, \aj, 107, 2240

\bibitem[Carpenter(2001)]{Carpenter2001} Carpenter, J.~M.\ 2001, \aj,
121, 2851

\bibitem[Carpenter(2003)]{Carpenter2003} Carpenter, J.~M.,\ 2003,
http://www.astro.caltech.edu/~jmc/2mass/v3/transformations/

\bibitem[Chanam{\'e} \& Gould(2004)]{Chaname2004} Chanam{\'e}, J., 
\& Gould, A.\ 2004, \apj, 601, 289

\bibitem[P. Dufour, P. Bergeron, G. Fontaine (2005)]{DBF05} P. Dufour,
P. Bergeron, G. Fontaine 2005, astro-ph/0503448 accepted for \apj

\bibitem[Davies et al.(2002)]{DKR02} Davies, M.~B., King, A., 
\& Ritter, H.\ 2002, \mnras, 333, 463

\bibitem[Digby et al.(2003)]{Digby2003} Digby, A.~P., Hambly, N.~C.,
Cooke, J.~A., Reid, I.~N., \& Cannon, R.~D.\ 2003, \mnras, 344, 583

\bibitem[Fontaine et al.(2001)]{FBB01} Fontaine, G., Brassard, P., \&
Bergeron, P.\ 2001, \pasp, 113, 409

\bibitem[Gilmore et al.(1989)]{GWK89} Gilmore, G., Wyse, R.~F.~G., \&
Kuijken, K.\ 1989, \araa, 27, 555

\bibitem[Gizis(1997)]{Gizis1997} Gizis, J.~E.\ 1997, \aj, 113, 806

\bibitem[Gizis \& Reid(1997)]{Gizis1997-2} Gizis, J.~E., \&
Reid, I.~N.\ 1997, \pasp, 109, 849

\bibitem[Gizis(1998)]{Gizis1998} Gizis, J.~E.\ 1998, \aj, 115, 
2053

\bibitem[Gizis \& Reid(1999)]{Gizis1999} Gizis, J.~E., \& Reid, I.~N.\
1999, \aj, 117, 508

\bibitem[Hansen(1999)]{H99} Hansen, B.~M.~S.\ 1999, \apj, 
520, 680

\bibitem[Hansen et al.(2002)]{Hansen2002} Hansen, B.~M.~S., et 
al.\ 2002, \apjl, 574, L155


\bibitem[Hawley et al.(1996)]{Hawley1996} Hawley, S.~L., Gizis, J.~E.,
\& Reid, I.~N.\ 1996, \aj, 112, 2799

\bibitem[Iben \& Laughlin(1989)]{IL89} Iben, I.~J., \& Laughlin, G.\
1989, \apj, 341, 312

\bibitem[Jao et al.(2003)]{Jao2003} Jao, W.-C., Henry, T.~J., 
Subasavage, J.~P., Bean, J.~L., Costa, E., Ianna, P.~A., \& M{\'e}ndez, 
R.~A.\ 2003, \aj, 125, 332 

\bibitem[Jao (2004)]{Jao2004} Jao, Wei-Chun, Ph.D. thesis, Georgia State
University

\bibitem[Jao et al.(2005)]{Jao2005} Jao, Wei-Chun, Henry, T.~J.,
Subasavage, J.~P., Brown, M.~A., Ianna, P.~A., Bartlett, J.~L., Costa,
E., \& M{\' e}ndez, R.~A.\ 2005, \aj, 129, 1954

\bibitem[Karaali et al.(2003)]{Karaali2003} Karaali, S., Bilir, S., 
Karata{\c s}, Y., \& Ak, S.~G.\ 2003, Publications of the Astronomical 
Society of Australia, 20, 165

\bibitem[Kleinman et al.(2004)]{Kle2004} Kleinman, S.~J., et al.\
2004, \apj, 607, 426

\bibitem[Leggett(1992)]{Leggett1992} Leggett, S.~K.\ 1992, \apjs, 
82, 351

\bibitem[L{\'e}pine \& Shara(2005)]{Lepine2005} L{\' e}pine, S., \&
Shara, M.~M.\ 2005, \aj, 129, 1483

\bibitem[Luyten (1939)]{Luyten1939} Luyten, W.~J., Bruce Proper Motion
Survey III. The Stars of Large Proper Motion and the Luminosity
Function, Minneapolis: University of Minnesota, 1939

\bibitem[Montgomery et al.(1999)]{MKWW99} Montgomery, M.~H., 
Klumpe, E.~W., Winget, D.~E., \& Wood, M.~A.\ 1999, \apj, 525, 482

\bibitem[Oppenheimer et al.(2001)]{Op01} Oppenheimer, B.~R., Hambly,
N.~C., Digby, A.~P., Hodgkin, S.~T., \& Saumon, D.\ 2001, Science,
292, 698

\bibitem[Reid et al.(2001)]{Reid2001} Reid, I.~N., Sahu, K.~C., \&
Hawley, S.~L.\ 2001, \apj, 559, 942

\bibitem[Reyl{\' e} et al.(2001)]{RRC01} Reyl{\' e}, C., 
Robin, A.~C., \& Cr{\' e}z{\' e}, M.\ 2001, \aap, 378, L53

\bibitem[Richer et al.(1997)]{Richer1997} Richer, H.~B., et al.\
1997, \apj, 484, 741 

\bibitem[Salim \& Gould(2003)]{Salim2003} Salim, S., \& Gould, A.\
2003, \apj, 582, 1011

\bibitem[Salim et al.(2004)]{Salim2004} Salim, S., Rich, R.~M.,
Hansen, B.~M., Koopmans, L.~V.~E., Oppenheimer, B.~R., \& Blandford,
R.~D.\ 2004, \apj, 601, 1075

\bibitem[Silvestri et al.(2002)]{Silvestri2002} Silvestri, N.~M.,
Oswalt, T.~D., \& Hawley, S.~L.\ 2002, \aj, 124, 1118

\bibitem[Silvestri et al.(2005)]{Silvestri2005} Silvestri, N.~M.,
Hawley, S.~L., \& Oswalt, T.~D.\ 2005, \aj, 129, 2428

\bibitem[Silvestri et al.(2002)]{SOH02} Silvestri, N.~M., 
Oswalt, T.~D., \& Hawley, S.~L.\ 2002, \aj, 124, 1118

\bibitem[Schmidt(1959)]{S59} Schmidt, M.\ 1959, \apj, 129, 243

\bibitem[van Altena et al.(1995)]{YPC} van Altena, W.~F., Lee, J.~T.,
\& Hoffleit, D.\ 1995, The General Catalogue of Trigonometric Stellar
Parallaxes (4th ed.; New Haven: Yale Univ. Obs.)

\bibitem[Weidemann(1987)]{W87} Weidemann, V.\ 1987, \aap, 188, 74

\bibitem[Weidemann(2000)]{W00} Weidemann, V.\ 2000, \aap, 
363, 647

\bibitem[Weidemann \& Koester(1983)]{WK83} Weidemann, V., \&
Koester, D.\ 1983, \aap, 121, 77

\bibitem[Weis(1996)]{Weis1996} Weis, E.~W.\ 1996, \aj, 112,
2300

\bibitem[West et al.(2004)]{West2004} West, A.~A., et al.\ 2004, \aj,
128, 426

\bibitem[Wood(1990)]{W90} Wood, M.~A.\ 1990, \jrasc, 84, 
150

\bibitem[Wood(1992)]{W92} Wood, M.~A.\ 1992, \apj, 386, 539

\bibitem[Wood(1995)]{W95} Wood, M.~A.\ 1995, LNP Vol.~443: 
White Dwarfs, 443, 41

\end{thebibliography}
\end{document}